\documentstyle[prb,aps]{revtex}
\begin{document}
\vspace*{0.7in}
\begin{center}
{\Large {\bf Kinetic Inductance and Penetration Depth of Thin
Superconducting Films Measured by THz Pulse Spectroscopy}}
\vskip .5in
S. D. Brorson$^{(a)}$, R. Buhleier, J. O. White$^{(b)}$,
I. E. Trofimov$^{(c)}$,
H. -U. Habermeier, and J. Kuhl \\
Max-Planck-Institut f\"ur Festk\"orperforschung, Stuttgart, Germany
\vskip .4in
PACS: 74.30.Gn, 74.70.Vy, 78.47.+p
\vskip 0.5in
{\bf Abstract}
\end{center}

We measure the transmission of THz pulses
through thin
films of YBa$_2$Cu$_3$O$_{7-\delta}$ at temperatures between
10 K and 300 K.  The pulses possess a usable bandwidth extending from
$\sim$0.1 -- 1.5~THz (3.3~cm$^{-1}$ -- 50~cm$^{-1}$).  Below $T_c$
we observe pulse reshaping caused by the kinetic inductance of
the superconducting charge carriers.  From transmission data, we extract
values of the London penetration depth, as a function of temperature,
and find that it agrees well with a functional form
$(\lambda(0)/\lambda(T))^2 = 1- (T/T_c)^{\alpha}$, where
$\lambda(0) = 148$~nm, and $\alpha = 2$.

\vskip 1in
\noindent
$^{(a)}$ also at Odense Universitet, DK-5230 Odense M, Denmark \\
$^{(b)}$ permanent address: Hughes Research Laboratories,
Malibu, CA, USA. \\
$^{(c)}$ permanent address: P. N. Lebedev Physics Institute, Moscow,
Russia.

\newpage

The unique electrodynamic properties of superconductors are a consequence
of the Meissner effect: magnetic fields attempting to penetrate a
superconductor are screened out by the flow of supercurrent.
The screening
takes place within a distance given by the London penetration depth,
$\lambda$.  It follows\cite{schrieffer} that the
frequency-dependent conductivity of a BCS--like superconductor
for frequencies far below the gap
takes the form
\begin{equation}
\label{sigma}
\sigma(\omega) = \sigma_1 + i \sigma_2 = \sigma_n + \left[\delta(\omega)
+ i/(\mu_o \omega \lambda^2) \right]
\end{equation}
where $\sigma_n$ is the contribution of the normal carriers, $\mu_0$
is the inductance of free space, and $i = \sqrt{-1}$.  The two terms
within square brackets represent the contribution of the superconducting
carriers.  At exactly $\omega = 0$, the conductivity is divergent, as
signified by the delta function $\delta(\omega)$.  For
$\omega \neq 0$, the superconducting part of $\sigma$ is purely imaginary,
and varies as $\omega^{-1}$.  Accordingly, the supercurrent responds to
the application of a time varying field like an ideal inductor, a property
referred to as ``kinetic inductance''.  The  magnitude of the kinetic
inductance depends upon $\lambda$.

Because of its importance for discriminating between different theories,
the temperature dependence of $\lambda$ in the high-$T_c$ materials has
become an object of active investigation.  Bonn, et al.\cite{bonn} have
recently pointed out that low frequency microwave (1 --
10~GHz)\cite{microwave_meas_2}, muon spin rotation\cite{bonn}, as well as
low-field magnetization measurements\cite{dcmag} often reveal a
temperature dependence of the form
\begin{equation}
\label{lambda}
(\lambda(0)/\lambda(T))^2 = 1 - (T/T_c)^\alpha
\end{equation}
where $\alpha = 2$.  This behavior is incompatible
with BCS theory\cite{bonn,BCS_alpha}, as well as the Gorter--Casimir
two-fluid model for which $\alpha = 4$ in the
neighborhood of $T_c$.\cite{2-fluid}  Contemporary theories positing
d-wave coupling for the Cooper pairs usually find $\alpha = 1$ in the $T
\rightarrow 0$ limit, although $\alpha = 2$ can be obtained if enough
scattering is built into the theory\cite{d-wave}.  Low frequency
microwave measurements showing $\alpha = 1$ for single crystals and
high-quality thin films below 20 K have been recently
published\cite{alpha_1_hardy,alpha_1_ma}.

In this paper, we describe transmission measurements on
thin films of YBa$_2$Cu$_3$O$_{7-\delta}$ (YBCO) in the
less-investigated {\em high frequency} range 0.1 -- 1.5~THz.  Thin
film samples were employed because suitable single crystals do not exist for
transmission measurements in this frequency regime.  Because our sample
quality is high, we do not anticipate problems associated with
defects, although effects associated with weak links at grain
boundaries cannot be ruled out.\cite{alpha_1_hardy,halbritter}  The
transmission measurements were made using wide-bandwidth electromagnetic
pulses generated and detected by microfabricated antenna
structures\cite{grischkowsky}.  The special  feature of this technique is
that it permits us to measure the  transmitted {\em electric field}, $E(t)$,
thereby enabling us to obtain the complex frequency domain conductivity
$\sigma(\omega)$, from which we  extract $\lambda$.  Similar experiments
have been previously performed to measure the so-called anomalous coherence
peak in $\sigma_1$ in YBCO\cite{nusskopf}.

The experiments are carried out using a THz spectrometer similar to that
described in Ref. \cite{grischkowsky}.
The system has a usable bandwidth extending from $\sim$0.1 -- 1.5~THz.
We use a 30~$\mu$m transmitting antenna fabricated on low
temperature grown GaAs \cite{LTGaAs}, biased to 50 V, and driven with
optical pulses from a
colliding pulse modelocked dye laser (CPM).
The receiver is a 30~$\mu$m antenna fabricated on ion-implanted
silicon-on-sapphire.
It was gated with a second pulse from the CPM laser, and the
photocurrent is detected with a current-sensitive lock-in amplifier.

The 155~nm YBCO film was grown on a 500~$\mu$m NdGaO$_3$ substrate by
laser ablation\cite{laser_ablation}.
We have found NdGaO$_3$ to be the ideal substrate for our purposes
because it remains transparent and nondispersive
over the entire spectral bandwidth of our pulses, as well as over the
entire range of temperatures investigated here.  The film was grown in
the (100) orientation, so that the polarization of our THz pulses was
parallel to the  ${\bf {\hat a}}$,${\bf {\hat b}}$ plane. The normal state
resistivity of the film, measured with a four-point probe,
 is linear in $T$, extrapolating to 0~$\mu \Omega$cm at $\sim$10 K.
The 100~$\mu \Omega$cm resistivity at 100 K indicates
high sample quality.  The DC superconducting transition is at
92 K, with a width of 0.5 K.

During the experiment, the sample is held in a continuous-flow He
cryostat which has been modified for transmission experiments
in the far-infrared by the installation of 2-cm thick
polyethylene windows. The temperature is stabilized
to better than 1 K.
The sample is mounted on a
Cu block having a 6~mm hole for transmission; a bare
substrate was mounted onto a similar hole on the same block 1~cm away.
The THz beam is focused onto the sample and
recollimated with a pair of plastic lenses.  Care is taken to
avoid leakage of THz radiation through the hole not in use.
At every temperature, a scan of the YBCO-film/substrate combination
is followed by a scan of the bare substrate.  The sample and reference
temporal scans are
Fourier transformed and divided, giving the complex
frequency-domain transmission of the YBCO film itself.

The time-domain transmission data shown in Fig.~1
show two noteworthy features.
First, the initial pulse is followed at regular intervals by multiple
reflections from within the NdGaO$_3$ substrate (inset to Fig.~1a).
The reflected pulses emerge without a (dispersive)
shape change, but with alternating signs, i.e. $+, -, +, -$, and so on.
The signs alternate because the NdGaO$_3$/vacuum interface is a dielectric
interface, which has a positive field reflection coefficient, $\Gamma > 0$,
while the YBCO/NdGaO$_3$ interface is a metallic (shorting) interface
for which $\Gamma\approx -1.$\cite{plasma_freq}

The second noteworthy feature of Fig.1 is the pulse reshaping that
occurs upon cooling the sample from 100 K to 85 K.  This reshaping, a direct
result of the superconductor's kinetic inductance, can be
understood with the aid of a transmission line analogy (Fig.~2).
If a transmission line is shorted in the middle by an inductor having an
impedance $Z = -i\omega L$, it acts as a
high-pass filter: the inductor is a short circuit for DC,
whereas extremely high frequencies pass by with undiminished
magnitude\cite{high_freq}.
The transfer function, $t(\omega)$, relating the input and output voltages,
$V_{i}(\omega)$ and $V_{o}(\omega)$, is
\begin{equation}
\label{trans_line}
t(\omega) = {V_o(\omega) \over V_i(\omega)} ={{2} \over {2 + Z_o/Z}}
\end{equation}
where $Z_o$ is the characteristic impedance of the transmission line.
This transfer function, with $Z_0$ set equal to the vacuum impedance, directly
corresponds to the complex field transmission coefficient measured by our
apparatus.

We have tested the applicability of the transmission line analog by
calculating the effect of this transfer function using a
measured THz pulse as the input $V_i(t)$.  In the THz regime, $Z_o \gg
Z$ giving $t(\omega) \propto -i\omega$.  Depicted in Fig.~3a is the
leading pulse transmitted through the  sample at 100 K (from Fig.~1a)
plotted on an expanded time scale. Because this pulse propagates through
normal state YBCO, it follows the same optical path as that in Fig.~1b,
but is not  modified by the superconducting transfer function
(\ref{trans_line}).  We Fourier transform this pulse,  multiply by
$-i\omega$, and inverse transform to obtain $V_o(t)$. This simulation
(Fig.~3b) agrees well with the experimentally obtained pulse which
propagates through the sample in its superconducting state (Fig.~3c).

Because we observe clearly the effects of the kinetic inductance, we are
able to determine $\lambda$ from our transmission
data. In the case of a superconducting film on a dielectric substrate of
index $n_s$, the transmission function is given by
\begin{equation}
\label{nusskopf}
t(\omega) = {{1+n_s} \over {1+n_s + Z_o/Z_{\rm eff} }}
\end{equation}
where $Z_{\rm eff}$ is the effective surface impedance of the
superconducting film \cite{nusskopf}.
Eqn.~(\ref{nusskopf}) is clearly similar to (\ref{trans_line}).
In general, the surface
impedance of an {\em infinite half-space} of a superconductor is $Z_s =
R_s + iX_s = \sqrt{-i\omega\mu_o/\sigma}$, where $\sigma$ is the complex
conductivity given by (\ref{sigma}) \cite{surface_cond}.  In the case of
{\em finite} film thickness, the effective surface reactance $X_{\rm eff}$
(the imaginary part of $Z_{\rm eff}$) must be corrected as $X_{\rm eff} =
X_s \coth(d/\lambda)$, where $d$ is the superconducting film thickness
\cite{Z_eff}. Recalling that $\sigma_1 \ll \sigma_2$ in the
superconducting state yields
\begin{equation}
\label{xeff}
X_{\rm eff} = \mu_o \omega \lambda \coth(d/\lambda)
\end{equation}
Thus, by measuring $t(\omega)$, we can determine $X_{\rm eff}$.  Then,
under the assumption that $\lambda$ remains constant over our frequency
range, we can invert (\ref{xeff}) and extract $\lambda$.

Using this procedure, we have obtained $\lambda$ values for a sequence
of temperatures ranging from 10 K up to $T_c$.  A least-squares fit of
$\lambda(T)$ to a linear function in $T$ for $T < 50$ K yields an
extrapolated $\lambda(0) = 148$~nm.
A plot of  $(\lambda(0)/\lambda(T))^2$ versus the reduced temperature
is shown in Fig.~4 taking the fitted $\lambda(0)$ and $T_c = 92$ K
determined from the DC resistivity measurements.
Comparing the data to theoretical curves of the functional
form (2) we find the best agreement for $\alpha = 2$
(Fig.~4, solid line).
Similar results have also
been obtained by several other investigators using other measurement
techniques as well as in other frequency
ranges\cite{microwave_meas_2,bonn,dcmag}.
Curves corresponding to the Gorter-Casimir two-fluid model \cite{2-fluid}
(for  which $\alpha = 4$), as well as standard BCS theory\cite{BCS_alpha}
are also shown (Fig.~4, dashed, dash-dotted lines).
The fit of the latter two curves to our experimental data seemes to be
significantly worse.

In conclusion, we observe the reshaping of THz pulses upon
transmission through thin films of superconducting YBCO.
The pulse reshaping is a direct time domain demonstration of the kinetic
inductance of the superconductor.
The reshaping may be described using a transmission line analog.
{}From measurements of the pulse transmission, we have extracted
values for the temperature-dependent penetration depth $\lambda(T)$.

We acknowledge the expert technical assistance
of S. Tippmann and A. Schulz.  SDB and JOW acknowledge support by the
Alexander von Humboldt Foundation.  JOW was supported by Hughes
Research Laboratories during part of this work.

\newpage

\begin{center}
{\bf Figure Captions}
\end{center}

\vskip .4in
\noindent
{\bf Fig. 1} Time dependent transmission of THz pulses through YBCO.
a) $T > T_c$ b) $T < T_c$, showing pulse reshaping.  In
both cases, trailing pulses with alternating signs resulting from
reflections at the NdGaO$_3$/YBa$_2$Cu$_3$O$_{7-\delta}$ interface are
visible (inset).

\vskip .3in
\noindent
{\bf Fig. 2} Transmission line analog of THz pulse experiment.  The
incident voltage pulse $V_i(t)$ enters from the left, propagates past
the shunt inductor $L$ which models the kinetic inductance of the
superconductor, and emerges on the right as $V_o(t)$, reshaped by the
transfer characteristic $t(\omega)$ from Eqn. (\ref{trans_line}).

\vskip .3in
\noindent
{\bf Fig. 3} a) The leading pulse from Fig. 1a on an expanded scale.
b) The calculated pulse shape when the pulse in a) is acted upon by
$t(\omega)$.
c) The measured leading pulse from Fig. 1b (expanded scale) for comparison.

\vskip .3in
\noindent
{\bf Fig. 4} London penetration depth $\lambda$ {\em vs.} $T$
determined from the transmission data via Eqns. (\ref{nusskopf}) and
(\ref{xeff}) (dots).  Also shown are theoretical curves of Eqn.
(\ref{lambda}) with $\alpha = 2$ (solid line), $\alpha = 4$ (dash-dotted
line), and BCS theory (dashed line).


\vskip .2in
\noindent
{\bf References}
\begin{thebibliography}{99}

\bibitem{schrieffer} J. R. Schrieffer, {\em Theory of Superconductivity}
(Benjamin/Cummings, Reading, MA, 1964).

\bibitem{bonn} D. A. Bonn, et al., Phys. Rev. B {\bf 47},
11314 (1993)

\bibitem{microwave_meas_2} S. M. Anlage, et al., Phys. Rev. B {\bf 44},
8764 (1991) (re-analyzed in \cite{bonn}); J. Lee and T. R. Lemberger,
Appl. Phys. Lett. {\bf 62}, 2419 (1993); M. R. Beasley, Physica C {\bf
209}, 43 (1993).

\bibitem{dcmag} A. M. Neminsky and P. N. Nikolaev, Physica C {\bf
212}, 389 (1993).

\bibitem{BCS_alpha} B. M{\"u}hlschlegel, Z. Phys. {\bf 155}, 313 (1959).

\bibitem{2-fluid}C. J. Gorter and H. G. B. Casimir, Phys. Z. {\bf 35},
963 (1934).

\bibitem{d-wave} J. Annett, N. Goldenfeld, and S. R. Renn, Phys. Rev.
B {\bf 43}, 2778 (1991); P. Arberg, M. Mansor, and J. P. Carbotte, Solid
State Comm. {\bf 86}, 671 (1993).

\bibitem{alpha_1_hardy} W. N. Hardy, et al., Phys. Rev. Lett. {\bf 70}, 3999
(1993).

\bibitem{alpha_1_ma} Z. Ma, et al., Phys. Rev. Lett. {\bf 71}, 781 (1993).

\bibitem{halbritter} J. Halbritter, J. Appl. Phys. {\bf 71}, 339 (1992).

\bibitem{grischkowsky} D. Grischkowsky, S. Keiding, M. van Exter, and C.
Fattinger, J. Opt. Soc. Am. B {\bf 7}, 2006 (1990).

\bibitem{nusskopf} M. C. Nuss, K. W. Goossen, P. M. Mankiewich, and M. L.
O'Malley, Appl. Phys. Lett. {\bf 58}, 2561 (1991).

\bibitem{LTGaAs} A. C. Campbell, G. E. Crook, T. J. Rogers, and B. G.
Streetman, J. Vac. Sci. Technol. B {\bf 8}, 305 (1990); M. Y. Frankel,
J. F. Whitaker, G. A. Mourou, F. W. Smith, and A. R. Calawa, IEEE Trans.
Electron Devices {\bf 37}, 2493 (1990).

\bibitem{laser_ablation} H. -U. Habermeier, G. Beddies, B. Leibold, G.
Lu, and G. Wagner, Physica C
{\bf 180}, 17 (1991); A. P. Litvinchuk, C. Thomsen, I. E. Trofimov, H. -U.
Habermeier, and M. Cardona, Phys. Rev. B {\bf 46}, 14017 (1992).

\bibitem{plasma_freq} The plasma frequency of YBCO is 120~THz.  See, for
example, D. E. Aspnes and M. K. Kelly, IEEE J. Quantum
Electron. {\bf 25}, 2378 (1989).

\bibitem{surface_cond} See, e.g., E. C. Jordan and K. G. Balmain, {\em
Electromagnetic Waves and Radiating Systems} (Prentice-Hall, Englewood
Cliffs, 1968), p. 153.

\bibitem{Z_eff} N. Klein, et al., J. Appl. Phys. {\bf 67}, 6940 (1990).

\bibitem{high_freq} Of course, in a real superconductor, loss caused by pair
breaking will occur once the photon energy becomes larger than the gap energy.

\end{thebibliography}
\end{document}